\documentclass[11pt,twoside]{article}

\usepackage{asp2014}

\aspSuppressVolSlug
\resetcounters

\begin{document}

\title{Implementation feedback of the IVOA Provenance data model}

\author{Mireille~Louys,$^{1,2}$ Fran\c cois~Bonnarel,$^2$ Daniel~Durand,$^3$ and Anais~Egner$^4$}
\affil{$^1$Universit\'e de Strasbourg, CNRS, ICube, UMR 7357, Strasbourg, France \email{mireille.louys@unistra.fr}}
\affil{$^2$Universit\'e de Strasbourg, Observatoire Astronomique de Strasbourg, CNRS UMR 7550, Strasbourg, France \email{francois.bonnarel@astro.unistra.fr}}
\affil{$^3$National Research Council Canada, Canadian Astronomy Data Centre, Victoria BC, Canada}
\affil{$^4$Universit\'e de Strasbourg, IUT Informatique, Illkirch, France}

\paperauthor{Louys Mireille}{mireille.louys@unistra.fr}{0000-0002-4334-1142}{University of Strasbourg}{ICube UMR7357-CNRS, Observatoire Astronomique UMR7357-CNRS} {Strasbourg}{}{67000}{France}
\paperauthor{Bonnarel Fran\c cois}{francois.bonnarel@astro.unistra.fr}{}{University of Strasbourg}{Observatoire Astronomique,  UMR7357-CNRS}{Strasbourg}{}{67000}{France}
\paperauthor{Durand Daniel}{daniel_durand@mac.com}{}{National Research Council Canada - Herzberg Astronomy and Astrophysics}{CADC}{Victoria}{BC}{}{Canada}
\paperauthor{Egner Anais}{anais.egner@live.fr}{}{Strasbourg University}{IUT Informatique, Illkirch}{Strasbourg}{}{67000}{France}


\begin{abstract}
The IVOA Provenance Data model defines entities, agents and activities as container classes to describe the provenance of datasets, with the executed tasks and responsibilities attached to agents. It also provides a set of classes to describe the activities type and their configuration template, as well as the configuration applied effectively during the execution of a task.
Here we highlight lessons learned in the implementation of the CDS ProvHiPS service distributing provenance metadata for the HST HiPS data collections, and for the HST archive original images used to produce the HiPS tiles.
ProvHiPS is based on the ProvTAP protocol, the emerging TAP standard for distributing provenance metadata. ProvTAP queries may rapidly become very complex.
Various graph representation strategies, including ad hoc solutions, triplestore and SQL CTE have been considered and are discussed shortly.
\end{abstract}

\section{Introduction: IVOA provenance data model and provenance of CDS HiPS}
Datasets used in astronomy are generally the results of a flow of observation and processing steps. Information on this process is generally called "provenance" of the dataset and is stored in various formats and logical organizations. This makes in general the provenance information difficult to compare and use interoperably among different data collections. That's the main reason why IVOA developed an astronomy-oriented provenance data model during the last years. This formalization not only allows traceability of products but also acknowledgment and contact information, quality and reliability assessment and discovery of datasets by provenance details.  At the time of writing the Provenance data model specification is an IVOA proposed recommendation \citep{2019ProvRec}.
HiPS \citep{fernique2015} defines a new way of organizing image, cube and catalogue data in an all sky and hierarchical way based on HealPIX tessellation of the sky. HiPS datasets are generated from image data collections or catalogues which have their own history. The ProvHiPS service developed at CDS aims at providing provenance information for HiPS stored at CDS back to original raw data when available.     

\section{HiPS datasets for HST image collections}
HiPS are made of hierarchies of tiles containing pixelized information at a given HealPIX order.
 In the case of image HiPS datasets each tile is generated from a small subset of the original image collection intersecting with the tile. The HiPS format stores inter-nally the progenitor information for each tile in the HiPS tree. The CDS data center publishes HiPS at various wavelengths  for the HST image collections.
 They have been produced from HST drizzled images in collaboration with CADC astronomer Daniel Durand. HST data collections are stored and retrievable from the CADC HST archive through IVOA DAL services.
 The drizzled images have their own history: they are produced from sets of calibrated images closely related on the sky by a specific type of co-addition called "drizzling". 

A rich tree of related data  is then potentially available. We have browsed HiPS tiles metadata and FITS headers of the HST images to extract features relevant in terms of the IVOA  Provenance data model to trace historical information  and map it into the ProvTAP tables. This resulted in a database containing ten of thousands of entities and activities. Dozen of descriptions of the various kinds of entities have also been produced as well as ten of thousands of configuration parameters for the drizzling and HiPS generation activities.   
     
\section{ProvHiPS implementation}
In order to make such information available we implemented a ProvTAP\footnote{https://wiki.ivoa.net/internal/IVOA/ObservationProvenanceDataModel/ProvTAP.pdf} service called ProvHiPS. ProvTAP specification is currently an IVOA working draft describing how to map the provenance data model in a TAP service. The heart of it is the Prov-TAP TAP SCHEMA definition providing the list of tables and columns required for storing the provenance metadata information and mapping respectively the classes and attributes of the model. Tables and columns come with datatype, unit, ucd, and utypes consistent with the model. 
By default a ProvTAP service is queriable via the ADQL language defined in IVOA and provides results in VOTable format.      

The CDS ProvHiPS service implements a database containing the provenance information sketched out in the workflow scenario presented in Fig \ref{fig:workflow}. 
\articlefigure[width=0.8\textwidth]{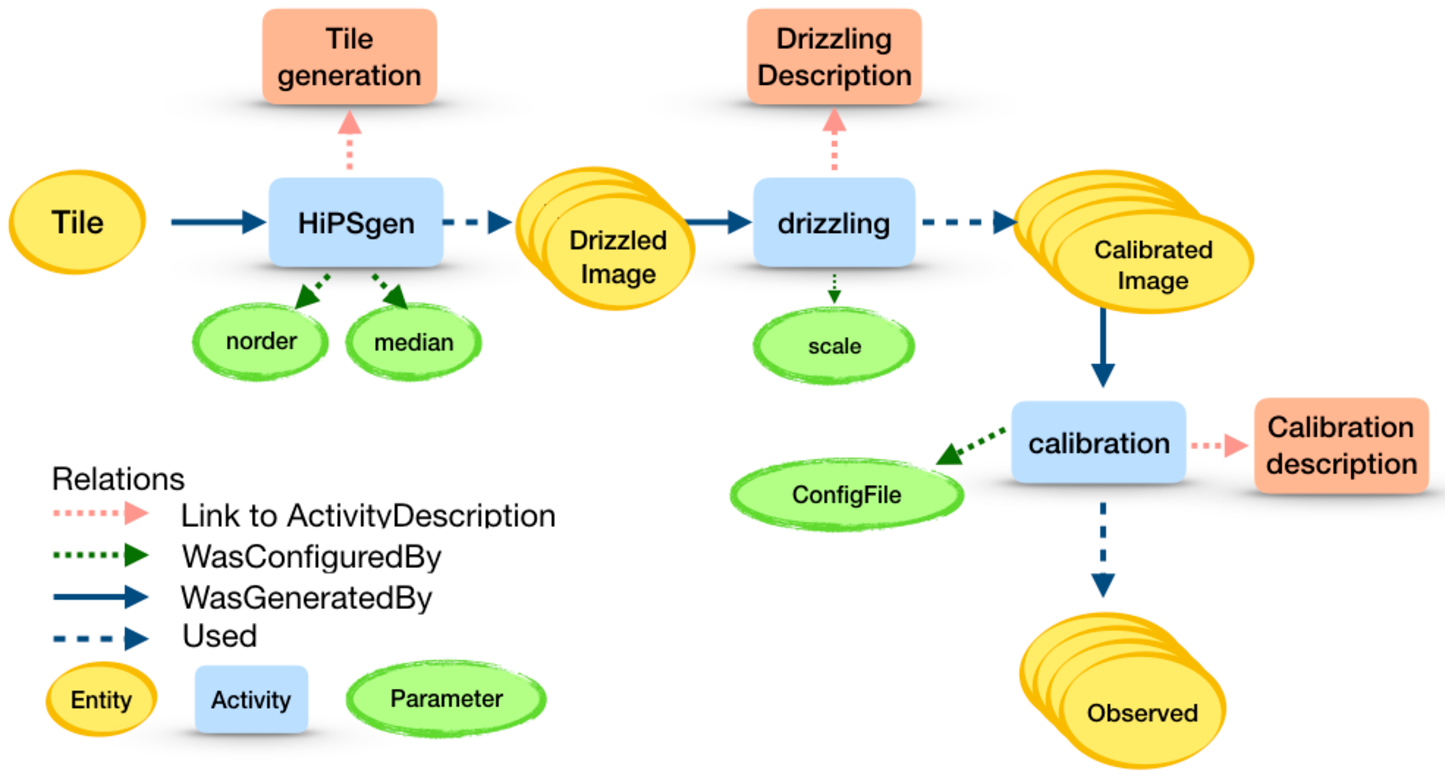}{fig:workflow}{The historical path from raw HST images to their projected tiles in the HiPS HST representation.}
\articlefigure[width=0.8\textwidth]{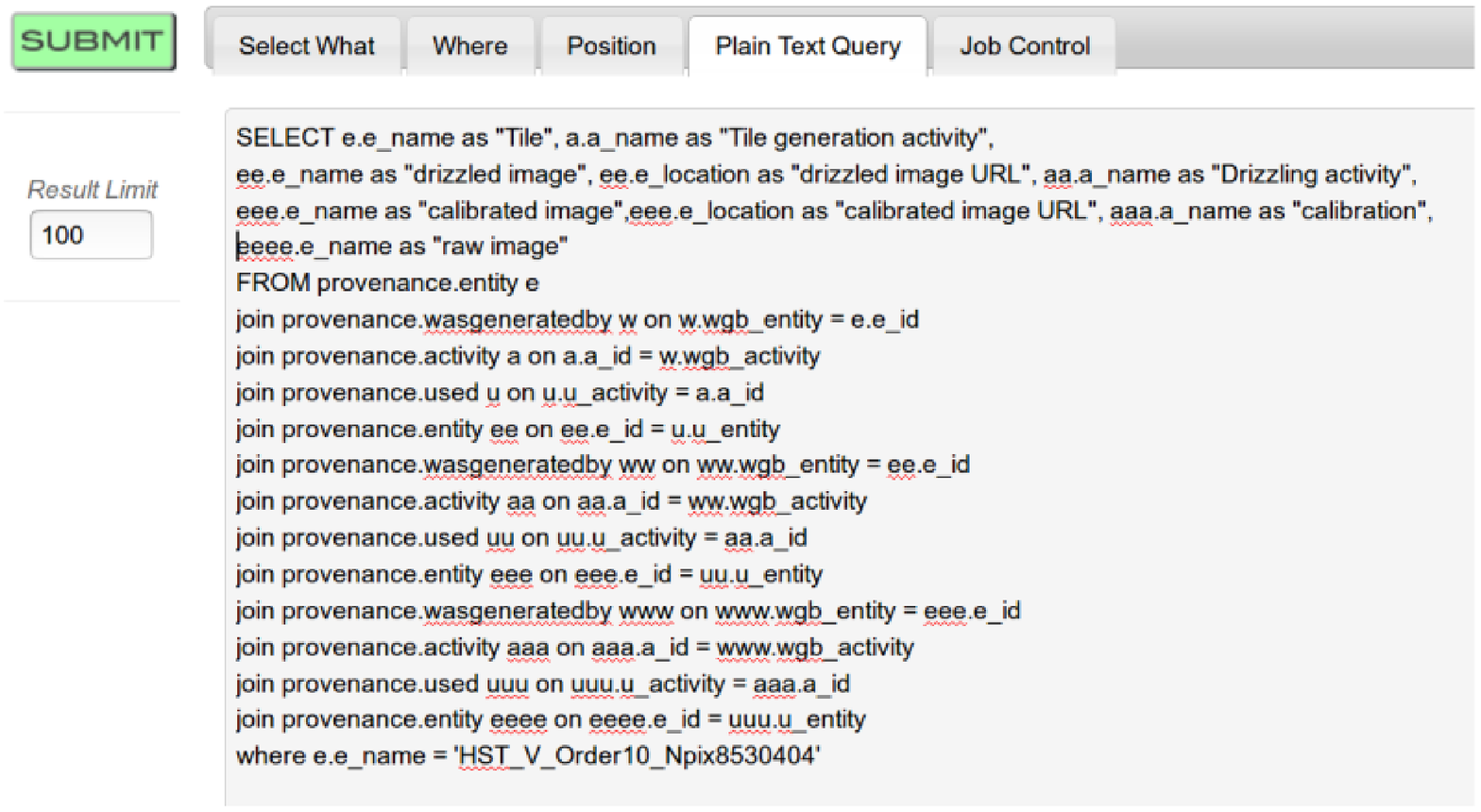}{fig:query}{Deep query from tile back to original raw image.}

Fig. \ref{fig:query} shows a 13-joins query tracing the provenance of a single HiPS V HST tile around the target NGC104. The query response is presented in Fig. \ref{fig:qresp} as displayed with the TAPHandle application \citep{2014ASPC..485...15M}. Some drizzled and calibrated images are visualized in Aladin via SAMP messaging as well.

\articlefigure[width=0.8\textwidth]{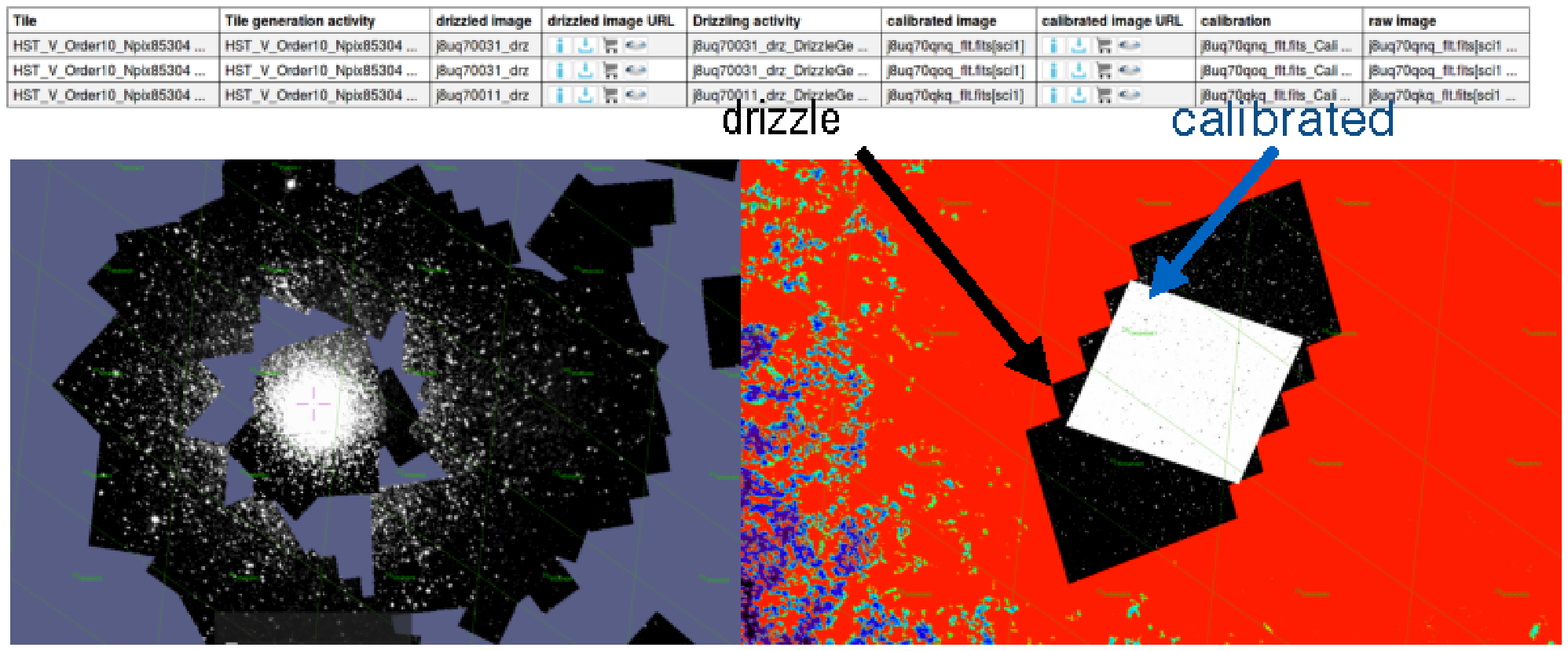}{fig:qresp}{Query response as shown in the TapHandle application. Access links to images at each step allow to understand the sequence of activities and how they transform data.}

 Fig. \ref{fig:desc} shows the activity description associated with one of the previous calibration activity. This Activity description provides interoperable typing of the calibration activities as well as a link to the software documentation. 
\articlefigure[width=0.8\textwidth]{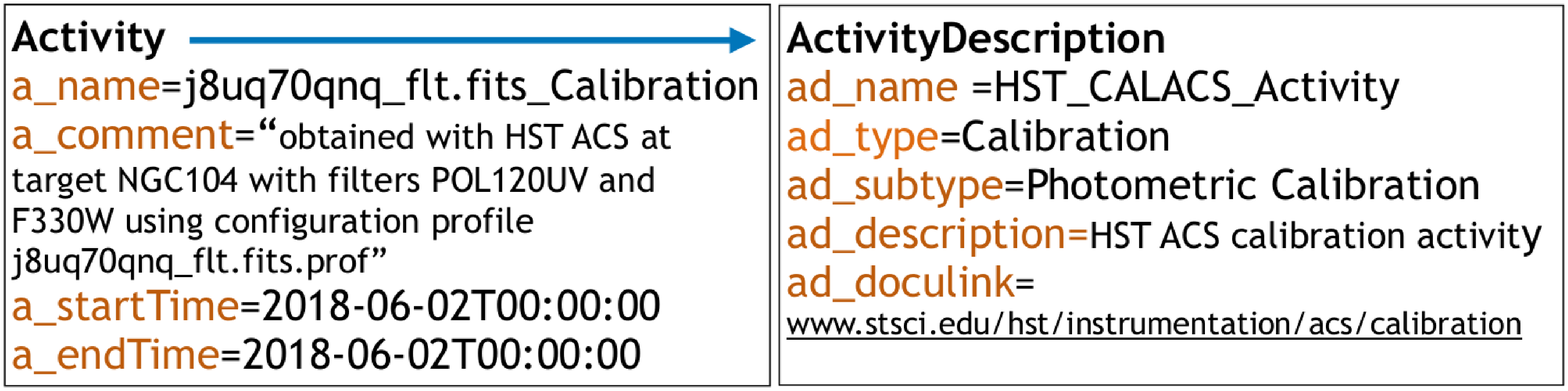}{fig:desc}{Metadata for an instance Activity for image calibration and its corresponding  ActivityDescription instance.}


More sophisticated user scenarios may include retrieving "siblings" of a given dataset entity using various depths or selecting datasets sharing the same creator agent or generated with similar parameters. The number of joins needed to traverse the provenance graph may increase tremendously. That's the reason why we experimented 
various ways of representing and querying graphs on top of relational databases. Former tests with a triplestore architecture has shown promising results \citep{2019ASPC..523...329}. As published in \citep{P2_15_adassxxix} and proposed by
M.Nullmeier\footnote{https://www.asterics2020.eu/dokuwiki/lib/exe/fetch.php?media=open:wp4:nullmeier\_tf5\_prov\_custom\_adql.pdf}, graphical or Common Table Expressions (CTE) techniques to navigate through graph connections on top of the RDBMS are  new solutions to consider. 
We plan to add such layers on top of our service to improve user-friendliness. 

\section{Conclusion}

Despite the "complex query" issue, the ProvTAP implementation of ProvHiPS demonstrates it is feasible to map information stored in FITS headers of homogeneous image collections or HiPS metadata into the IVOA PROV DM profile. The scalability of the database allows coping with very large data collections. Retrieval of multiple steps pipelines is easy as long as the  appropriate ADQL queries are provided.

\acknowledgements We thank the CDS intership program for supporting A. Egner. This work has been partly supported by the ESCAPE project (the European Science Cluster of Astronomy and Particle Physics ESFRI Research Infrastructures) funded by the EU Horizon 2020 research and innovation program under the Grant Agreement n.824064, and also the ASTERICS project under Grant Agreement n.653477.

\bibliography{P2-6}  


\end{document}